\documentclass[showpacs,prl,twocolumn,floatfix,
nofootinbib,superscriptaddress]{revtex4}
\usepackage{graphicx}

\newcommand{\be}{\begin{equation}}
\newcommand{\ee}{\end{equation}}
\newcommand{\nl}{\nonumber \\}
\newcommand{\msb}{{\overline{\rm MS}}}

\begin{document}


\title{ $B^0_s-\overline{B^0_s}$ 
Mixing Parameters from Unquenched  Lattice QCD}

\author{Emel Dalgic}
\affiliation{Department of Physics,
The Ohio State University, Columbus, OH 43210, USA}
\author{Alan Gray}
\affiliation{Department of Physics,
The Ohio State University, Columbus, OH 43210, USA}
\author{Elvira Gamiz}
\affiliation{Department of Physics \& Astronomy,
University of Glasgow, Glasgow, G12 8QQ, UK}
\author{Christine T.\ H.\ Davies}
\affiliation{Department of Physics \& Astronomy,
University of Glasgow, Glasgow, G12 8QQ, UK}
\author{G.\ Peter Lepage}
\affiliation{Laboratory of Elementary Particle Physics,
Cornell University, Ithaca, NY 14853, USA}
\author{Junko Shigemitsu}
\affiliation{Department of Physics,
The Ohio State University, Columbus, OH 43210, USA}
\author{Howard Trottier}
\affiliation{Physics Department, Simon Fraser University,
Vancouver, British Columbia, Canada}
\author{Matthew Wingate}
\affiliation{Institute for Nuclear Theory, University of Washington,
Seattle, WA 98195-1550, USA}

\collaboration{HPQCD Collaboration}
\noaffiliation


\begin{abstract}
We determine hadronic matrix elements relevant for the mass and width 
differences, $\Delta M_s$ \& $\Delta \Gamma_s$, in the $B^0_s-\overline{B^0_s}$
 meson system using fully unquenched lattice QCD.
We employ the MILC collaboration gauge configurations that include $u$, 
$d$ and $s$ sea quarks using the improved staggered quark (AsqTad) 
action and a highly improved gluon action. We implement the valence 
$s$ quark also with the AsqTad action and use NonRelativistic QCD for 
the valence $b$ quark. 
For the nonperturbative QCD input into the 
Standard Model expression for $\Delta M_s$ we find
$f_{B_s} \, \sqrt{\hat{B}_{B_s}} = 0.281(21)$GeV. 
 Results for 
 four-fermion operator matrix elements entering 
Standard Model formulas for $\Delta \Gamma_s$ are also presented.
\end{abstract}


\maketitle


\section{Introduction}

Recent developments at the Tevatron Run II have dramatically improved 
our knowledge of the mass difference $\Delta M_s$ between the ``heavy'' 
and ``light'' mass eigenstates in the $B_s^0 - \overline{B^0_s}$ system. The 
Spring of 2006 witnessed first the two-sided bound on $\Delta M_s$ by 
the D\O $\,$ collaboration \cite{d0} followed quickly by a precise  
measurement of this quantity by the CDF collaboration \cite{cdf}.
 $B_s$ mixing occurs in the Standard Model through box diagrams 
with two $W$-boson exchanges. These diagrams can be reexpressed 
in terms of an effective Hamiltonian involving four-fermion 
operators. In order to compare the Tevatron measurements with 
Standard Model predictions, matrix elements of the four-fermion 
operators between the $B^0_s$ and $\overline{B^0_s}$ states must be 
computed. Only then can one test for consistency between 
experiment and the Standard Model and, in the case that precise agreement fails
 to be realized,  hope to discover hints of new physics.
 $B_s^0 - \overline{B^0_s}$ mixing is a $\Delta B = 2$ process and 
 sensitive to effects of physics beyond the Standard Model. Hence 
a large effort is underway to nail down the Standard Model predictions 
as accurately as possible. 
In the current article we present a fully unquenched lattice QCD determination 
of the hadronic matrix elements of several crucial four-fermion 
operators. 

\section{Simulation Details}

Our simulations use the MILC collaboration $N_f=2+1$ 
unquenched gauge configurations \cite{milc1}. To date we have completed 
calculations on two of the MILC coarse ensembles with the 
light sea quark mass $m_f$ satisfying 
 $m_f/m_s = 0.25$ and $m_f/m_s=0.5$ respectively and with $m_s$ being 
the physical strange quark mass. 
For the strange valence quark we use the improved staggered 
(AsqTad) \cite{stagg} quark action.  
The $b$-quark is simulated using the same improved 
nonrelativistic (NRQCD) action employed  in recent studies of 
the $\Upsilon$ system \cite{upsilon}
 and for calculations of the $B$ and $B_s$ meson 
decay constants \cite{fbprl,fbsprl} 
and the $B \rightarrow \pi, l \overline{\nu}$ 
semileptonic form factors \cite{bsemi}. 
As in our previous work using the MILC configurations we use the 
$\Upsilon$ 2S-1S splitting to fix the lattice spacing, which in the present 
case gives  $a^{-1} = 1.596(30)$GeV and $a^{-1} = 1.605(29)$GeV \cite{upsilon}
for the $m_f/m_s=0.25$ and $m_f/m_s=0.5$ ensembles respectively. 
The bare $b$ and $s$ quark masses have likewise  been 
fixed already in previous simulations of the $\Upsilon$ \cite{upsilon}
and kaon \cite{milc2} systems.
Some theoretical issues remain having to do with the need 
to take a fourth root of the AsqTad action determinant while 
creating the MILC unquenched configurations. This procedure is 
the focus of intense scrutiny by the lattice community and there has been 
considerable progress in our understanding of the issues involved 
during the past year\cite{fourthroot}. To date 
 no obstacles have been found to invalidate obtaining true QCD in the 
continuum limit.
 The MILC configurations and the light and heavy 
quark actions employed in this article have also been tested  
by comparing the results of accurate calculations for a large range of 
hadronic quantities to experimental results \cite{prl,mbc,dsemi,fds,alpha}.  
 The outcome of these 
tests have been very encouraging.  Here we apply the same
 successful lattice approach to $B^0_s - \overline{B^0_s}$ mixing.

\section{The Four-Fermion Operators and Matching}

We have studied the following four-fermion operators that enter into 
calculations of 
$\Delta M_s$ and $\Delta \Gamma_s$ in the Standard Model 
(``i'' and ``j'' are color indices)
\begin{eqnarray}
\label{fourfol}
 OL & \equiv & [\overline{b^i} \, s^i]_{V-A} [\overline{b^j} \,
 s^j]_{V-A} ,\\
\label{fourfos}
 OS  &\equiv&  [\overline{b^i} \, s^i]_{S-P} [\overline{b^j}\,
 s^j]_{S-P} ,\\
\label{fourfo3}
 O3  &\equiv&  [\overline{b^i} \, s^j]_{S-P}
[\overline{b^j}\, s^i]_{S-P} .
\end{eqnarray}
In continuum QCD in the $\overline{MS}$ scheme matrix elements 
of these operators are parametrized in terms of the $B_s$ meson 
decay constant $f_{B_s}$ and so-called ``bag'' parameters $B(\mu)$ 
at some scale $\mu$, 
\be
\label{defol}
\langle OL \rangle ^{\overline{MS}}_{(\mu)}
\equiv
\langle \overline{B}_s | OL| B_s \rangle ^{\overline{MS}}_{(\mu)} 
\equiv \frac{8}{3}
f^2_{B_s} \, B_{B_s}(\mu)\,  M^2_{B_s}.
\ee
The factor $\frac{8}{3}$ is inserted so that $B_{B_s}=1$ corresponds 
to the ``vacuum saturation'' approximation. The 
four-fermion operators $OS$ and $O3$ have similarly
 each their own bag parameter
\be
\label{defos}
\langle OS \rangle ^{\overline{MS}}_{(\mu)}
\equiv 
 - \frac{5}{3}
f^2_{B_s} \,\frac{ B_S(\mu)}{R^2}\,  M^2_{B_s},
\ee
\be
\label{defo3}
\langle O  3 \rangle ^{\overline{MS}}_{(\mu)}
\equiv 
 \frac{1}{3}
f^2_{B_s} \, \frac{\tilde{B}_S(\mu)}{R^2}\,  M^2_{B_s},
\ee
with
\be
\frac{1}{R^2} \equiv \frac{M_{B_s}^2}{(\overline{m}_b + \overline{m}_s)^2}.
\ee
The Standard Model expression for the mass difference $\Delta M_s$ 
is given by \cite{buras},
\be
\label{deltams}
\Delta M_s = \frac {G_F^2 M_W^2}{6 \pi^2} |V^*_{ts}V_{tb}|^2 \eta_2^B
S_0(x_t) M_{B_s} f^2_{B_s} \hat{B}_{B_s},
\ee
where $x_t = m_t^2/M_W^2$,  $\eta_2^B$ is  a perturbative QCD correction
factor,  $S_0(x_t)$ the Inami-Lim function and $V_{ts}$ and $V_{tb}$
 the appropriate Cabibbo-Kobayashi-Maskawa (CKM) matrix elements. 
  The nonperturbative QCD 
input into this formula is the combination $f^2_{B_s} \, \hat{B}_{B_s}$ with 
$\hat{B}_{B_s}$ the renormalization group invariant bag parameter. 
At two-loops and using 
 $n_f=5$ and $\; \alpha^{(n_f=5)}_{\overline{MS}}(\mu=m_b=4.8{\rm GeV})
 = 0.212$ \cite{alpha} one finds $\hat{B}_{B_s}/B_{B_s} = 1.534$.

 In order to evaluate hadronic matrix elements of the four-fermion 
operators via lattice QCD methods, one must first relate the 
operators in continuum QCD to operators written in terms of lattice 
heavy and light quark fields.  We carry out this matching between 
continuum QCD and the lattice theory through ${\cal O}
(\alpha_s)$, ${\cal O}(\frac{\Lambda_{QCD}}{M})$ and ${\cal O}
(\frac{\alpha_s}{aM})$.  Our lattice theory works 
with NRQCD $b$-quarks.  At lowest order in $1/M$ the $b$ fields 
in (\ref{fourfol}) - (\ref{fourfo3}) must be replaced by NRQCD 
heavy quark or heavy anti-quark fields.  The tree-level relation 
between NRQCD and full QCD fields is given by the Foldy-Wouthuysen-Tani 
transformation. At ${\cal O}(\frac{\Lambda_{QCD}}{M})$ this 
brings in dimension seven corrections to the four-fermion operators, 
which are of the form,
\begin{eqnarray}
\label{fourfj1op}
  OLj1   
&\equiv & \frac{1}{2M} \left\{ [\vec{\nabla}\overline{b^i}
 \cdot \vec{\gamma} \, s^i]_{V-A} [\overline{b^j} \,
s^j]_{V-A}  \right.  \nl
 &&   \left. \quad + \quad
  [\overline{b^i} \, s^i]_{V-A} [\vec{\nabla}\overline{b^j}
 \cdot \vec{\gamma} \, s^j]_{V-A}  \right\}  .
\end{eqnarray}
Similar $1/M$ corrections $OSj1$ and $O3j1$ can be introduced 
for the four-fermion operators $OS$ and $O3$.
To the order stated above, matching between $\langle OX \rangle^\msb$ 
($X$ = $L$,$S$ or $3$) 
and matrix elements in the lattice theory is then given by (we suppress 
the $\mu$ dependence),
\begin{eqnarray}
\label{ox}
&&\frac{a^3}{2 M_{B_s} } \; \langle OX \rangle ^{\msb} =  \nl
&&[ 1 + \alpha_s \cdot \rho_{XX} ] \langle OX \rangle
+ \alpha_s \cdot \rho_{XY} \langle OY \rangle \; +  \nl
&& \left [ \langle OXj1 \rangle - \alpha_s ( \zeta^{XX}_{10}
\langle OX \rangle + \zeta^{XY}_{10} \langle OY \rangle)
\right ] .
\end{eqnarray}
 $\langle OX \rangle$ without the superscript $\msb$ 
 stands for the matrix element in the lattice theory.
Even at lowest order in $1/M$ there is mixing 
between the four-fermion operators. 
At ${\cal O}(\alpha_s)$ 
the mixing occurs between 
$X,Y=L$ and $S$ for $\langle OL\rangle$ 
and  $\langle OS\rangle$ and between 
$X,Y=3$ and $L$ for $\langle O3\rangle$. 
This mixing  takes place already in continuum QCD when one carries 
out an expansion in $1/M$ \cite{flynn,hashimoto}. 
Due to the good chiral properties of AsqTad light quarks, which we use 
for the  valence $s$ quark in our simulations, no additional operator 
mixing arises upon going from the continuum to the lattice theory. 
We have multiplied $\langle OX \rangle^\msb$ in (\ref{ox}) 
  by a factor of $\frac{a^3}{2M_{B_s}}$ in order to 
take into account the different normalization of states in QCD and 
the lattice theory and also to render the lattice matrix elements 
$\langle OX \rangle$ dimensionless. 
Details of calculations of the one-loop coefficients $\rho_{XY}$ 
and $\zeta^{XY}_{10}$ will be presented in a separate paper. The methodology 
is similar to that of \cite{pert1,pert2}.  As in those matching calculations 
for heavy-light currents, the 
$\alpha_s \cdot \zeta^{XX}_{10}$ and 
$\alpha_s \cdot \zeta^{XY}_{10}$ terms 
in (\ref{ox}) are necessary to subtract ${\cal O}(
\frac{\alpha_s}{aM})$ power law contributions from 
the matrix elements $\langle OXj1 \rangle$.

\section{Simulation Results and Error Estimates}

The hadronic matrix elements $\langle \hat{O} \rangle$, 
$\hat{O}$=OX and OXj1, are determined by evaluating three-point 
correlators via numerical simulations, 
\be
\label{thrpnt}
  C^{(4f)}(t_1,t_2) = 
 \sum_{\vec{x}_1,\vec{x}_2} \langle 0 |
\Phi_{\overline{B}_s}(\vec{x}_1,t_1) \; [\hat{O} ](0) 
\; \Phi^\dagger_{B_s}(\vec{x}_2,-t_2) | 0 \rangle.
\ee
 $\Phi_{B_s}$ is an interpolating operator for the $B_s$ meson and the 
four-fermion operator $\hat{O}$ is fixed at the origin of the lattice. 
We fit $C^{(4f)}$ together with the $B_s$ meson two-point correlator,
 $C^B(t)$, to the following forms
\begin{eqnarray}
\label{cf4fform}
 && C^{(4f)}(t_1,t_2) =  \nl
&& \sum_{j,k=0}^{N_{exp}-1}  A_{jk} \;
(-1)^{j \cdot t_1} \; (-1)^{k \cdot t_2} \; e^{-E_B^{(j)} (t_1-1)}
\; e^{-E_B^{(k)}(t_2-1)}, \nl
\end{eqnarray}
\begin{eqnarray}
C^B(t) &=& \sum_{\vec{x}} \langle 0| \Phi_{B_s}(\vec{x},t)
\; \Phi^\dagger_{B_s}(0) | 0 \rangle \nl
& = &\sum_{j=0}^{N_{exp}-1}
 \xi_j \; (-1)^{j \cdot t} \; e^{-E_B^{(j)}(t-1)} .
\end{eqnarray}
The dimensionless
 matrix elements entering the right-hand side (RHS)
 of (\ref{ox}) are then given by,
\be
 \langle \hat{O} \rangle 
= \frac{A_{00}}{\xi_0}.
\ee
Results for $\langle \hat{O} \rangle$ are summarized in Table I 
for our two dynamical ensembles.  The errors are 
combined statistical and fitting uncertainty errors. More details on 
our fits are given in \cite{lat06}. 
In Table I we also show results for $\langle OXj1 \rangle_{(sub)}$, 
the true relativistic corrections (after power law subtractions)
 from the dimension seven operators. 
By considering 
$\langle OXj1 \rangle_{(sub)}/\langle OX \rangle$, one finds
the physical ${\cal O}(\Lambda_{QCD}/M)$
contribution to be -13\% for $\langle OL \rangle$, 
11\% for $\langle OS \rangle$ and  6$\sim$8\% for $\langle O3 \rangle$.

\begin{table}
\caption{Matrix elements in the lattice theory for fixed 
$b$ and $s$ valence masses and two values 
of the light $u$, $d$ sea quark mass. Errors are statistical plus fitting 
errors. }
\begin{center}
\begin{tabular}{|c|cc|}
\hline
  & $\;\;m_f/m_s = 0.25\;\;$  & $\;\;m_f/m_s = 0.50\;\;$ \\
\hline
$\langle OL \rangle$ & 0.1036(83)    & 0.1069(92)\\
$\langle OS \rangle$ & -0.0680(54)    & -0.0687(61)\\
$\langle O3 \rangle$ & 0.0143(12)    & 0.0142(13)\\
\hline
$\langle OLj1 \rangle$ & -0.0227(18)    & -0.0229(18)\\
$\langle OSj1 \rangle$ & -0.0130(10)    & -0.0139(11)\\
$\langle O3j1 \rangle$ & 0.0021(3)    & 0.0026(3)\\
\hline
$\langle OLj1 \rangle_{(sub)}$ & -0.0138(20)    & -0.0140(20)\\
$\langle OSj1 \rangle_{(sub)}$ & -0.0072(11)    & -0.0081(12)\\
$\langle O3j1 \rangle_{(sub)}$ & 0.0008(3)    & 0.0012(4)\\
\hline
\end{tabular}
\end{center}
\end{table}

Having determined the matrix elements in the lattice theory 
we can plug the numbers into the RHS of (\ref{ox}).  For this 
matching we use 
$\alpha_s = \alpha_V^{(n_f=3)}(2/a)=0.32$ \cite{alpha}.
 We set the scale for $\alpha_s$ to $q^* = 2/a$,  which is close 
to $q^*$'s evaluated for heavy-light currents using other 
heavy and light quark actions.
The matching coefficients $\rho_{XY}$ are generally functions of 
the $\msb$ scale $\mu$ through the combination $\log(\frac{\mu}{m_b})$.
 We present results for $\mu = m_b$.  We evaluate the RHS of 
(\ref{ox}) for 
$\langle OL \rangle^{\msb}$, 
$\langle OS \rangle^{\msb}$ and 
$\langle O3 \rangle^{\msb}$ and combine with the definitions in 
(\ref{defol}) - (\ref{defo3}) to obtain the main results of this article, 
namely, 
\be
\label{mainres}
f^2_{B_s} \, B_{B_s}, \;\;\;\;
f^2_{B_s} \, \frac{B_S}{R^2}, \;\;\;\;
f^2_{B_s} \, \frac{\tilde{B}_S}{R^2}.
\ee
The main errors in these quantities are listed in Table II. 
One sees that the two dominant errors are due to statistics + fitting and 
higher order matching uncertainties. We have also included a nonnegligible 
error coming from the uncertainty in the scale (lattice spacing) for the 
MILC ensembles used.  At the final stage of 
extracting  results for (\ref{mainres}), one has to convert $a^3 f^2_{B_s} 
\, M_{B_s}$ into physical units. An uncertainty of $\sim 1.8$\% 
in the lattice spacing turns at this point into a $\sim 5$\% uncertainty for 
$a^{-3}$. The leading discretization error in the actions employed here 
comes in at ${\cal O}(a^2 \alpha_s) \sim 2$\% and is believed to be 
dominated by taste-changing effects in the Asqtad action, an assumption 
that has been checked recently by comparing Asqtad valence quarks with more 
highly improved staggered quarks from the HISQ action \cite{hisq}.
  We multiply 
the $2$\% by a factor of 2 to come up with a total discretization 
uncertainty of $4$\%, which should cover taste-changing effects in the sea 
as well, including the fourth root.
  This total error is consistent with scaling tests carried
  out via explicit simulations at two lattice spacings of other 
  B physics quantities such as decay constants and
  semileptonic form factors employing the same actions as in
  the present article \cite{fbprl,bsemi}.
  In the future we plan to repeat 
the current calculations at finer lattice spacings in order to reduce 
discretization uncertainties and also to carry out tests with HISQ 
instead of AsqTad light valence quarks.

In Table II we take the operator matching 
 error to be $ 1 \times \alpha_s^2$ since matching is done directly for 
the combination $f^2_{B_s} \, B_{B_s}$ (and for the other quantities 
in (\ref{mainres})). A naive attempt to deal separately with 
$f^2_{B_s}$ and $B_{B_s}$ in the formula for $\Delta M_s$ could 
increase the error estimate since the perturbative error for 
just $f_{B_s}$ alone (unsquared) is usually also taken as 
$1 \times \alpha_s^2$ coming from higher order 
matching of the heavy-light current.
We avoid unnecessarily separating out the bag parameters and possibly 
introducing  ambiguities in error 
estimates by always working with the relevant combination 
$f^2_{B_s} \, B_{B_s}$.
  Several years ago reference \cite{soni} also
 emphasized the virtues of working with physical combinations and 
never splitting off the bag parameters. The possibility of reducing 
errors by focusing on the combined $f^2_{B_q}B_{B_q}$ is also mentioned 
in \cite{damir}.

\begin{table}
\caption{Error budget for quantities listed in (\ref{mainres}).
}
\begin{center}
\begin{tabular}{|lc|}
\hline
Statistical + Fitting  &  9 \% \\
Higher Order Matching  &  9 \% \\
Discretization         &  4 \% \\
Relativistic           &  3 \% \\
Scale ($a^{-3}$)       &  5 \% \\
\hline
 Total        &   15 \%\\
\hline
\end{tabular}
\end{center}
\end{table}

\begin{table}
\caption{Results for the square root of quantities listed in
 (\ref{mainres}).
The unhatted bag parameters are given at scale $\mu = m_b$.
Errors quoted are combined statistical and systematic errors.
Note that percentage errors here are smaller than those given in Table II
by a factor of two due to the square root.}
\begin{center}
\begin{tabular}{|c|cc|}
\hline
  & $\;\;m_f/m_s = 0.25\;\;$  & $\;\;m_f/m_s = 0.50\;\;$ \\
\hline
&& \\
$f_{B_s} \sqrt{\hat{B}_{B_s}}$ [GeV] & 0.281(21)    & 0.289(22)\\
&& \\
$f_{B_s} \sqrt{B_{B_s}(m_b)}$ [GeV] & 0.227(17)    & 0.233(17) \\
\hline
&& \\
$f_{B_s} \frac {\sqrt{B_S(m_b)}}{R}$  [GeV] & 0.295(22) & 0.301(23)  \\
&& \\
$f_{B_s} \frac {\sqrt{\tilde{B}_S(m_b)}}{R}$  [GeV] & 0.305(23) &0.310(23)  \\
\hline
\end{tabular}
\end{center}
\end{table}

Table III gives our final values for the square root of the 
quantities listed in (\ref{mainres}) 
together with the scale invariant combination
 $f_{B_s} \sqrt{\hat{B}_{B_s}}$.
One sees that the light sea quark mass dependence is 
 small and not statistically significant. 
The largest difference between the central values of the 
 $m_f/m_s=0.25$ and $m_f/m_s = 0.5$ 
results is less than $3$\%, smaller than any of the other 
errors, and in particular significantly  smaller than our current 
statistical errors. Any reasonable estimate of 
chiral extrapolation (in $m_{sea}$) uncertainties will not 
affect the total error in Table II.
In the future we plan to use  Staggered Chiral 
perturbation theory (SChPT) \cite{schpt1,schpt2}
 to extrapolate to the physical chiral 
limit.  SChPT formulas for $B_q$ mixing with Asqtad light quarks 
are being worked out by  Laiho and Van de Water \cite{ruth} 
and will be important in $B_d$ mixing studies where one 
needs to extrapolate in both the valence and sea light quark masses.
In the present case of $B_s$ mixing and until our statistical errors 
have been further reduced and more data points are available, 
we do not believe the whole machinery of SChPT is crucial.
 We note that in our studies of the $B_s$ meson 
decay constant, where we have data for $f_{B_s}$ at four different 
light sea quark masses on coarse MILC ensembles and for two sea quark 
masses on the fine MILC lattices, no sea quark mass dependence was 
observed \cite{fbprl,lat05js}. 
We do not attempt a chiral extrapolation in the light sea 
quark mass with the current data and take the 
$m_f/m_s = 0.25$ numbers as our best determinations of the hadronic 
matrix elements.
In particular, this gives for the combination 
$f_{B_s}\sqrt{ \hat{B}_{B_s}}$, the crucial nonperturbative ingredient 
for $\Delta M_s$, the value quoted in the abstract:
\be
\label{result}
f_{B_s} \, \sqrt{\hat{B}_{B_s}} = 0.281(21) \, {\rm GeV}.
\ee

\section{Results for the Mass Difference $\Delta M_s$ and for 
$|V_{ts}^*V_{tb}|$} 

Using (\ref{result}) one can now attempt a theory prediction for 
$\Delta M_s$ based on the Standard Model.  We plug in standard values for the 
other ingredients in (\ref{deltams}) taken from recent reviews. 
We use $\eta_2^B = 0.551(7)$, $\overline{m}_t(m_t)=162.3(2.2)$GeV 
(which leads to $S_0(x_t)=2.29(5)$) and 
$|V_{ts}^* V_{tb}| = 4.1(1) \times
10^{-2}$ \cite{ckm,ball} together with (\ref{result}) to obtain 
\be
\Delta M_s (theory) = 20.3(3.0)(0.8) ps^{-1}.
\ee
The first error is the $15$\% error from $f^2_{B_s} \,\hat{B}_{B_s}$ and 
the second an estimate of the error from $|V_{ts}^*V_{tb}|$ and 
$\overline{m}_t$. 
$\Delta M_s (theory)$ is consistent with the CDF measurement of 
$17.77\,\pm 0.10\,\pm 0.07\, ps^{-1}$ \cite{cdf}.
 At the moment theory errors, 
in other words the lattice errors, dominate.  However, one is 
already in a position 
 to place nontrivial constraints on beyond the Standard 
Model effects.  An alternative way to test consistency of the Standard Model 
is to use the CDF measurement of $\Delta M_s$ together with (\ref{result}) 
to determine 
$|V_{ts}^* V_{tb}|$.  One finds 
\be
\label{vts}
|V_{ts}^* V_{tb}| = 3.9(3) \times 10^{-2},
\ee
 consistent with the value used above which is based on the measured 
 value for $|V_{cb}|$ plus unitarity \cite{ball}.  
The error in (\ref{vts}) comes entirely from the uncertainty in
$f_{B_s} \sqrt{\hat{B}_{B_s}}$. 

\section{Results for the Width Difference $\Delta \Gamma_s$ }

We have emphasized our result (\ref{result}) since this enters into 
$\Delta M_s$ for which a precision experimental measurement exists. 
The other entries in Table III for the four-fermion operators 
``OS'' and ``O3'' are relevant for the width difference $\Delta \Gamma_s$ 
\cite{beneke,lenz}, for which experimental errors are currently still
greater than $50$\%.  A recent measurement by the D\O $\,$ Collaboration 
\cite{d0dgs} gives, for instance, $\Delta \Gamma_s = [0.13 \pm 0.09] ps^{-1}$.
On the theory side, the authors of \cite{lenz} have recently shown that 
 by going to a new basis employing operators ``OL'' \& ``O3'',
as opposed to the old basis of ``OL'' \& ``OS'', theoretical uncertainties 
from $1/m_b$ and $\alpha_s$ corrections can be significantly reduced.
We insert our results from Table III for $f_{B_s} \sqrt{B_{B_s}}$ and 
$f_{B_s} \frac{\sqrt{\tilde{B}_S}}{R}$, taking again the $m_f/m_s = 0.25$ 
data, into eq.(51) of reference \cite{lenz} and obtain,
\be
\label{delgs}
\Delta \Gamma_s =  0.10(3) \, ps^{-1},
\ee
which is consistent with the D\O $\,$ measurement. About half the error 
in (\ref{delgs}) comes 
from lattice errors in $f^2_{B_s} B_{B_s}$ and $f^2_{B_s} \frac{\tilde{B}_S}
{R^2}$ and the other  half from remaining theoretical uncertainties in
the formula of \cite{lenz}. Both types of errors can be reduced through 
further work on the lattice and we plan to focus on obtaining a more accurate 
Standard Model theory prediction for $\Delta \Gamma_s$ while waiting for 
the experimental measurements to improve as well.

We argued above that there is no need to separate out the bag parameters 
when calculating $\Delta M_s$ or $\Delta \Gamma_s$.  
One might nevertheless be interested in doing so to 
 judge how close we are to the ``vacuum saturation'' 
approximation.  To convert to bag parameters we use the central value of the 
$B_s$ meson decay constant determined in \cite{fbsprl}, 
$f_{B_s}=0.260(29)$GeV, together with (for $1/R^2$) \cite{lenz}
$\overline{m}_b = 4.25$GeV and $\overline{m}_s = 85$MeV. 
One finds $B_{B_s}(m_b) = 0.76(11)$, $B_S(m_b) = 0.84(13)$ and 
$\tilde{B}_S(m_b) = 0.90(14)$, values consistent with 
earlier quenched \cite{hashimoto,becirevic} and $N_f=2$ \cite{jlqcd} 
 lattice determinations.

\section{Summary}
This article presents full QCD 
 results for hadronic matrix 
elements of four-fermion operators relevant for $B^0_s - 
\overline{B^0_s}$ mixing. We give unquenched results including the 
effect of $2+1$ flavors of sea quarks and calculate both leading and 
next-to-leading matrix elements in a nonrelativistic expansion 
of four-fermion operators for the first time.
Using our nonperturbative QCD results one 
finds agreement between the recent Tevatron 
measurements of the mass and width differences
 $\Delta M_s$ \& $\Delta \Gamma_s$ and Standard Model 
predictions. 
Our dominant errors for $f^2_{B_s} \, \hat{B}_{B_s}$ 
come from statistics + fitting 
and from higher order operator matching uncertainties. Work is 
underway aimed at reducing these errors. This will allow even tighter 
 constraints on any beyond the Standard Model 
effects entering $B^0_s - \overline{B^0_s}$ mixing phenomena. 
We have also initiated studies of  $B^0_d - \overline{B^0_d}$ mixing. 
 Our goal there is to obtain precision results for the important 
ratio $f^2_{B_s} B_{B_s} / f^2_{B_d} B_{B_d}$, which will provide further 
consistency checks on the Standard Model and  give us 
a handle on the CKM matrix element $|V_{td}|$. Some errors listed 
 in Table II will cancel almost completely in this ratio, such as the 
$a^{-3}$ and the higher order matching uncertainties.  Others, such as the 
statistical and fitting error, will cancel partially. 

\vspace{.1in}
{\bf Acknowledgements}: \\
This work was supported by the DOE and NSF (USA) and by PPARC (UK).
Numerical work was carried out at NERSC. We thank the MILC collaboration 
for making their unquenched gauge configurations available to us.



\end{document}